\documentclass[aps,prl,twocolumn,amsmath,amssymb,nofootinbib,superscriptaddress]{revtex4-1}

\usepackage{graphicx}
\usepackage{dcolumn}
\usepackage{bm}
\usepackage{amsmath}
\usepackage{indentfirst}
\usepackage{float}
\usepackage[colorlinks]{hyperref}
\usepackage[dvipsnames]{xcolor}
\usepackage{braket}
\usepackage{amsthm}
\begin{document}

\title{Demonstration of the Essentiality of Entanglement in a Deutsch-like Quantum Algorithm}

\author{He-Liang Huang}
\email[]{quanhhl@mail.ustc.edu.cn}
\affiliation{CAS Centre for Excellence and Synergetic Innovation Centre in Quantum Information and Quantum Physics,\\
University of Science and Technology of China, Hefei, Anhui 230026, China}
\affiliation{Hefei National Laboratory for Physical Sciences at Microscale and Department of Modern Physics, University of Science and Technology of
China, Hefei, Anhui 230026, China}

\author{Ashutosh K. Goswami}
\email[]{ashutoshgoswami841@gmail.com}
\affiliation{Indian Institute of Science Education and Research Kolkata, Mohanpur 741 246, West Bengal, India}

\author{Wan-Su Bao}
\email[]{glhhl0773@126.com}
\affiliation{CAS Centre for Excellence and Synergetic Innovation Centre in Quantum Information and Quantum Physics,\\
University of Science and Technology of China, Hefei, Anhui 230026, China}

\author{Prasanta K. Panigrahi}
\email[]{pprasanta@iiserkol.ac.in}
\affiliation{Indian Institute of Science Education and Research Kolkata, Mohanpur 741 246, West Bengal, India}
\date{\today}

\pacs{03.67.Ac, 03.65.Ud, 03.67.Lx, 42.50.Dv}

\begin{abstract}

Quantum algorithms could efficiently solve certain classically intractable problems by exploiting quantum parallelism. To date, whether the quantum entanglement is useful or not for quantum computing is still a question of debate. Here, we present a new quantum algorithm to show that entanglement could help to gain advantage over classical algorithm and even the quantum algorithm without entanglement. Furthermore, we implement experiments to demonstrate our proposed algorithm using superconducting qubits. Our results show the viability of the algorithm and suggest that entanglement is essential in getting quantum speedup for certain problems in quantum computing, which provide a reliable and clear guidance for developing useful quantum algorithms in future.
\newline Keywords: quantum computing; quantum entanglement; quantum algorithm; Deutsch's problem.
\newline PACS indexing codes: 03.67.Ac, 03.67.Lx, 03.65.Ud 

\end{abstract}

\maketitle
\section{Introduction}
Quantum information has undergone a revolutionary change in recent years. In 1982, the legendary physicist, R. P. Feynman noted that simulating $n$ qubits on a classical computer needs exponential resources, as it requires storing and processing of $2^n$ complex amplitudes \cite{Feynman}. However, a quantum computer based on the laws of quantum physics can naturally simulate $n$ qubits. This attractive advantage has driven the field of quantum computing. To date, considerable effort has gone into realizing the dream of practical quantum computers \cite{Monz,Wangboson,huangboson,Heboson,huanggate,zumetrology,Zhang2017,Nigg,Wang,Huang,Huangibmcloud,huangblind,Lu,Wu,Bernien2017,Barends2016,Barends2014}.

Harnessing the intrinsic nature of quantum mechanics, quantum superposition principle, quantum computers promise to give rise to an exponential speedup over their classical counterparts for certain tasks \cite{Nielsen,Ladd}. As the core for the speedup in quantum computing, quantum algorithms run on a realistic model of quantum computing. Design of well-performing quantum algorithms for important problems has been an interesting intellectual challenge and achievement all along. Notable examples include Shor's algorithm \cite{Shor}, Grover/Long algorithm \cite{Grover,Long}, Simon's algorithm \cite{Simon}, quantum simulation \cite{Feynman,Lloyd1996}, solving linear systems \cite{Harrow}, and quantum machine learning \cite{Rebentrost,Lloyd2014}. To get more valuable suggestions and experience for the design of quantum algorithms, it is of great importance to investigate the quantum-mechanical effects in the quantum algorithm, especially the role of different types of quantum resources in quantum algorithms.

Entanglement, a specific and magical type of quantum superposition, is the  quantum property of multiparticle systems that can not be written as a tensor product of individual quantum states. Entanglement has been used as a useful quantum resource in several quantum cryptographic and communication tasks \cite{Ekurt,Bennett,Wiesner}. However, its role in getting quantum speedup has not been established yet.

It has been shown that several quantum algorithms such as Bernstein-Vazirani \cite{Vazirani} and Grover search  \cite{Grover} do not require entanglement for their implementation \cite{Llyod,Meyer}. Biham $et$ $al.$ \cite{Biham} have shown that certain advantages of quantum algorithms remain even in the absence of entanglement.  Biham $et$ $al.$ \cite{Biham2002}  have studied how well a state performs as an input to Grover's search algorithm, and they found that the more the entanglement in the input, the less well the algorithm performs. Although some works have been proposed to studied the mechanism of quantum speedup \cite{Castagnoli}, and many \cite{Mor, Jozsa1, Ding} have argued that entanglement is necessary for quantum algorithms, not any specific example is provided as a clear evident to show the role of entanglement.

Here, we provide a quantum algorithm extending the Deutsch problem \cite {Deutsch,Jozsa} for two black boxes of two functions, which rely on entanglement for quantum speedup in an essential manner. To show the role of entanglement, we point out that a classical algorithm or a quantum algorithm without entanglement needs at least three queries to functions. However, only two queries are required in the proposed quantum algorithm with entanglement. Furthermore, a proof-of-principle demonstration is reported to show the viability of the proposed algorithm. For the first time, our work clearly demonstrates that the entanglement is the essentiality for quantum speedup of certain problems.

\section{Theory}

Before introducing our proposed algorithm, we first briefly describe the Deutsch's algorithm \cite{Deutsch}. Given a black box executing certain unknown function $f:\big\{0, 1\big\} \rightarrow \big\{0, 1\big\},$ one wishes to know whether the function $f$ is constant ($f(0)\oplus f(1)=0$) or balanced ($f(0)\oplus f(1)=1$). Classically, one needs two queries to the function $f$ to solve this problem, while Deutsch's algorithm can solve the problem in only a single query as follows (See Fig. 1):

\begin{itemize}
    \item Initializing two qubits to $\Ket{0}_{a_1}\otimes\Ket{1}_{a_2}$ and applying a Hadamard gate to each qubit. This yields
    $$ \frac{\Ket{0}_{a_1}+\Ket{1}_{a_1}}{\sqrt2} \otimes \frac{\Ket{0}_{a_2}-\Ket{1}_{a_2}}{\sqrt2}$$
     \item Applying the function $f$ to the current state, then we obtain the following state,
       $$ \frac{\Ket{0}_{a_1}+(-1)^{f(0)\oplus f(1)}\Ket{1}_{a_1}}{\sqrt2} \otimes \frac{\Ket{0}_{a_2}-\Ket{1}_{a_2}}{\sqrt2}$$
    \item Applying Hadamard gate and subsequently, measuring the qubit $a_1$ in computational basis.
\end{itemize}

\begin{figure}
\centering
\includegraphics[width=0.8\linewidth]{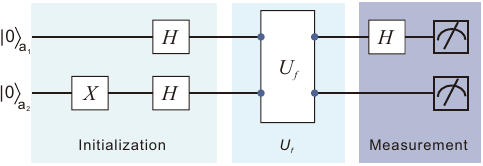}
\caption{(color online). Circuit for Deustch's Algorithm. The information that function is constant or balanced is stored in first qubit. The blue dots represent the qubits involved in the calculation of black box ($U_f$ or $U_g$).}
\end{figure}

Obviously, $f(0)\oplus f(1)=0$ if and only if we measure a zero and $f(0)\oplus f(1)=1$ if and only if we measure a one. So with certainty we could decide whether the function is constant or balanced. Evidently, Deutsch's algorithm has solved this problem in only one query, which is faster than classical algorithm.

However, we note that the Deutsch's algorithm could not be used to prove the advantage of entanglement, since no entanglement is generated in the algorithm. Next, we will propose a modified problem, and then prove that that entanglement could help to gain advantage over classical algorithm and even the quantum algorithm without entanglement for solving this specific problem.

Now, let us consider a similar problem. Assume that Alice has black boxes of the two unknown functions $f:\big\{0, 1\big\} \rightarrow \big\{0, 1\big\} $
and $g:\big\{0, 1\big\} \rightarrow \big\{0, 1\big\} $. She has been assured that both functions $f$ and $g$  are either constant or balanced, that is, $f(0) \oplus f(1) = g(0) \oplus g(1)$. Alice wants to compute following two quantities with minimum possible queries to the functions $f$ and $g$,

\begin{itemize}
\item $f(0) \oplus f(1)$ or $g(0) \oplus g(1)$; functions $f$ and $g$ are constant or balanced.
\item $f(0) \oplus g(0)$ or $f(1) \oplus g(1)$; functions $f$ and $g$ are same or different.
\end{itemize}

It is clear that classically, we need two queries to the function $f$ (or $g$) to compute $f(0)$ and $f(1)$ (or $g(0)$ and $g(1)$) and one query to the function $g$ (or $f$) to compute $g(0)$ (or $f(0)$).

Here, we propose a quantum algorithm exploiting quantum entanglement, which requires only one query to the each function $f$ and $g$. Thus, the proposed algorithm saves one query compared to the classical one. Subsequently, we point out that this quantum advantage is not possible without entanglement. Following is the step by step presentation of the proposed algorithm (see Fig. 2):

\begin{figure}[htbp]
\centering
\includegraphics[width=\linewidth]{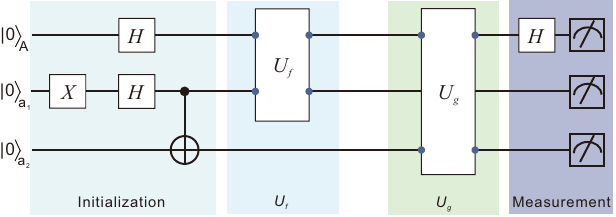}
\caption{(color online). Circuit for the proposed algorithm with entanglement. The information that functions are constant or balanced is stored in the first qubit, while information that they are same or different is stored in the second and third qubits. $U_f$ is applied on the 1-st and 2-nd qubits, and $U_g$ is applied on the 1-st and 3-rd qubits.}
\end{figure}

\begin{itemize}
    \item We start with one qubit $\Ket{0}_A$ and two ancilla qubits $\Ket{0}_{a_1}$ and $\Ket{0}_{a_2}$, and initialize these qubits to
    $$\frac{\Ket{0}_A+\Ket{1}_A}{\sqrt2} \otimes \frac{\Ket{0}_{a_1}\Ket{0}_{a_2}-\Ket{1}_{a_1}\Ket{1}_{a_2}}{\sqrt2}$$
    In the initialization step, two ancilla qubits are entangled.
    \item Quantum state of the composite system after applying functions $f$ and $g$ (ignoring normalization coefficients) is given by,
    $$ \Ket{0}_A( \Ket{0\oplus f(0)}_{a_1}\Ket{0\oplus g(0)}_{a_2}-\Ket{1\oplus f(0)}_{a_1}\Ket{1\oplus g(0)}_{a_2})$$  $$ + \Ket{1}_A (\Ket{0\oplus f(1)}_{a_1}\Ket{0\oplus g(1)}_{a_2}-\Ket{1\oplus f(1)}_{a_1}\Ket{1\oplus g(1)}_{a_2}) $$
     \textbf{Case -1}: If $f(0) \oplus g(0) =f(1) \oplus g(1)= 0$, that is, $f(0)=g(0)$ and $f(1)=g(1)$, the above equation reads,
     $$\Ket{0}_A (-1)^{f(0)} (\Ket{0}_{a_1}\Ket{0}_{a_2}-\Ket{1}_{a_1}\Ket{1}_{a_2})$$ $$+ \Ket{1}_A (-1)^{f(1)}(\Ket{0}_{a_1}\Ket{0}_{a_2}-\Ket{1}_{a_1}\Ket{1}_{a_2})$$
     which is equivalent to,
    $$ (\Ket{0}_A + (-1)^{f(0)\oplus f(1)}\Ket{1}_A)(\Ket{0}_{a_1}\Ket{0}_{a_2}-\Ket{1}_{a_1}\Ket{1}_{a_2})$$
     \textbf{Case -2}: For $f(0) \oplus g(0) = f(1) \oplus g(1)= 1$, that is, $f(0)\neq g(0)$ and $f(1)\neq g(1)$, it  reads,
     $$\Ket{0}_A (-1)^{f(0)} (\Ket{0}_{a_1}\Ket{1}_{a_2}-\Ket{1}_{a_1}\Ket{0}_{a_2})$$ $$+ \Ket{1}_A (-1)^{f(1)}(\Ket{0}_{a_1}\Ket{1}_{a_2}-\Ket{1}_{a_1}\Ket{0}_{a_2}),$$ which is equivalent to,
      $$ (\Ket{0}_A + (-1)^{f(0)\oplus f(1)}\Ket{1}_A)(\Ket{0}_{a_1}\Ket{1}_{a_2}-\Ket{1}_{a_1}\Ket{0}_{a_2})$$
    \item Applying Hadamard gate on qubit $A$, and then measuring the three qubits in computational basis.
\end{itemize}

It is obvious that the measurement outcome of qubit $A$ determines whether the functions $f$ and $g$ are constant or balanced, while the measurement outcomes of the two ancilla qubits determines whether they are same or different as depicted in Table I.

\begin{table}[htbp]
\centering
\begin{tabular}{|c |c|}
\hline
 \shortstack{measurement outcome (first qubit A)} & $f(0) \oplus f(1)$ \\
\hline
$\Ket{0}_A$ & 0 \\
\hline
$\Ket{1}_A $& 1 \\
\toprule
\shortstack{ measurement outcome (ancilla qubits)} & $f(0) \oplus g(0)$	\\
\hline
\shortstack{$\Ket{0}_{a_1}\Ket{0}_{a_2}$ or $ \Ket{1}_{a_1}\Ket{1}_{a_2}$}  & 0 \\
\hline
\shortstack{$\Ket{0}_{a_1}\Ket{1}_{a_2}$ or $\Ket{1}_{a_1}\Ket{0}_{a_2}$}  & 1 \\
\hline
\end{tabular}
\caption{The measurement results of the proposed algorithm, which determine the properties of functions $f$ and $g$.}
\end{table}


In the proposed algorithm above, entanglement is generated during the computing, and only two queries are required. In the computational complexity theory, it is customary to analyze algorithms with respect to the number of queries. This method of analyzing algorithms is called the query model. In the query model, an algorithm is said to be more efficient if it queries the oracle less number of times \cite{Ciliberto}. Next, we will prove that at least three queries are required if we can't generate any entanglement during the computing.

\textit{Theorem.} It is impossible to compute $f(0) \oplus f(1)$ and $f(0) \oplus g(0)$ together in overall two queries if we don't use quantum entanglement in the algorithm.

\textit{Proof.} Here, we assume that it is possible to compute $f(0) \oplus f(1)$ and $f(0) \oplus g(0)$ together in one query to each function $f$ and $g,$ without generating entanglement in any intermediate stage of the algorithm and derive a contradiction. In this case, the structure of the algorithm can be concluded as following,

\begin{itemize}
\item Initial state is $\Ket{\psi_0} \Ket{\psi_1} \Ket{\psi_2}$.
\item After applying the function $f,$ the state of the composite system $0,$ $1$ and $2$ changes into,
$$U_{f_{01}}(\Ket{\psi_0} \Ket{\psi_1} \Ket{\psi_2}) = \Ket{\phi}_{01} \Ket{\psi_2}$$
\item One can  perform some local quantum operations changing state $\Ket{\phi}_{01}$ into $\Ket{\xi}_{01}$ after applying $f.$
The state of the composite system $0,$ $1$ and $2$ after applying function $g$ is given by,
$$U_{g_{02}}(\Ket{\xi}_{01} \Ket{\psi_2})$$
\end{itemize}

Let's consider the second step of algorithm: Since it has been assumed that algorithm does not utilize entanglement, $\Ket{\phi}_{01}$ must be a product state,
$$U_{f_{01}}(\Ket{\psi_0} \Ket{\psi_1})=\Ket{\phi}_{01} = \ket{\chi_0} \otimes \Ket{\chi_1}$$

If function $f$ is constant, it is easy to see that $\Ket{\phi}_{01}$ will always be a product state. Thus, above relation will be true for any initial state $\Ket{\psi_0} \Ket{\psi_1}.$ However, problem arises when function $f$ is balanced. Without loss of generality, let's take $f(0)=0$ and $f(1)=1.$ In this case unitary $U_{f_{01}}$ is controlled-NOT (CNOT$_{01}).$ If $\Ket{\psi_0} = \alpha \Ket{0} + \beta \Ket{1}$, $\Ket{\psi_1} = \gamma \Ket{0} + \delta \Ket{1}$ then,
$$U_{f_{01}}(\Ket{\psi_0} \Ket{\psi_1}) = \alpha \gamma \Ket{00} + \alpha \delta \Ket{01} +\beta \delta \Ket{10}+  \beta \gamma \Ket{11} $$

For it to be a product state, $\alpha \beta (\gamma^2 - \delta^2) = 0 $ $\Rightarrow \alpha = 0$ or, $\beta = 0 $ or, $\gamma= \pm \delta.$ Thus, possible states that do not generate entanglement in this case read,
\begin{itemize}
\item $\Ket{0} (\gamma \Ket{0} + \delta \Ket{1})$
\item $\Ket{1} (\gamma \Ket{0} + \delta \Ket{1})$
\item $(\alpha \Ket{0} + \beta \Ket{1}) (\frac{\Ket{0}+\Ket{1}}{\sqrt2})$
\item $(\alpha \Ket{0} + \beta \Ket{1}) (\frac{\Ket{0}-\Ket{1}}{\sqrt2})$
\end{itemize}

Now we will see what information can we compute using these states,
\begin{itemize}
\item If state is of type  $\Ket{0} (\gamma \Ket{0} + \delta \Ket{1})$ or $\Ket{1} (\gamma \Ket{0} + \delta \Ket{1}), $ that is, the corresponding output after applying $U_{f_{01}}$ is $\Ket{0} (\gamma \Ket{0+f(0)} + \delta \Ket{1+f(0)})$ or $\Ket{0} (\gamma \Ket{0+f(1)} + \delta \Ket{1+f(1)}).$ At best, we can learn the value of $f(0)$ or $f(1),$ when either $\gamma=0$ or, $\delta=0$ is true.

\item If state is of type $(\alpha \Ket{0} + \beta \Ket{1}) (\frac{\Ket{0}+\Ket{1}}{\sqrt2}),$ after applying $U_{f_{01}}$ output will be,

$$ \alpha \Ket{0} (\frac{\Ket{0+f(0)}+\Ket{1+f(0)}}{\sqrt2}) + \beta \Ket{1} (\frac{\Ket{0+f(1)}+\Ket{1+f(1)}}{\sqrt2}).$$
This would always be $(\alpha \Ket{0} + \beta \Ket{1}) (\frac{\Ket{0}+\Ket{1})}{\sqrt2})$ independent of mapping $f.$ Since input and output is same, no information about the function $f$ is obtained after the computation.

\item  If state is of type $(\alpha \Ket{0} + \beta \Ket{1}) (\frac{\Ket{0}-\Ket{1}}{\sqrt2}),$ after applying $U_{f_{01}}$ output will be, $$(\alpha \Ket{0} + (-1)^{f(0)+f(1)}\beta \Ket{1}) (\frac{\Ket{0}-\Ket{1}}{\sqrt2}).$$
At best, one can learn $f(0)+f(1),$ when $\alpha  = \beta = \frac{1}{\sqrt2}.$
\end{itemize}

Thus, we can do only one computation out of  $f(0),$ $f(1)$  and $f(0) \oplus f(1)$  at best in one query without generating entanglement in the second step of algorithm. After execution of function $f$, the 0-th and 1-st qubits are in certain product state $\ket{\chi_0} \Ket{\chi_1}$,  and one can apply a local unitary on state $\ket{\chi_0} \otimes \Ket{\chi_1},$ changing it into $\ket{\xi_0} \otimes \Ket{\xi_1}$ before applying the function $g.$ However, after applying $U_{g_{02}}(\Ket{\xi}_{01} \Ket{\psi_2})$ and again demanding that the output should be a product state, similar to the second step we can do only one computation out of $g(0),$ $g(1),$ and $g(0) \oplus g(1)$  at best in one query.

Hence, in two queries, one can compute only one quantity from each set $A =\{ f(0), f(1), f(0) \oplus f(1)\}$ and $B=\{ g(0), g(1), g(0) \oplus g(1)\},$ without generating entanglement in any stage of the algorithm. However, no combination $(x, y),$ where $x\in A,$ $y\in B$ gives us $f(0) \oplus f(1)$ and $f(0) \oplus g(0)$ together, thus we derived a contradiction and this completes the proof.\qed

Therefore, when restricted to only one query to each of the functions $f$ and $g$, it is not possible to calculate logical quantities $f(0) \oplus f(1) $ and $f(0) \oplus g(0)$ together (given that both functions are either constant or balanced), using a classical computer or using a quantum computer without entanglement. Figure~3 is a example of quantum algorithm without entanglement which need three queries. By implementing the algorithm in Fig.~3, we can determine that the functions $f$ and $g$ are constant (balanced) if the measurement result of first qubit is ${\rm{|0}}\rangle$ (${\rm{|1}}\rangle$), and the functions $f$ and $g$ are same (different) if the measurement results of second and third qubits are same (different). In fact, the measurement results of the algorithm in Fig.~3 is the same to the Table I. We note that  both the proposed algorithms in Fig.~2 and Fig.~3 are deterministic. However, the algorithm in Fig.~2 only needs two queries, and the algorithm in Fig.~3 needs three queries. Thus, we have shown that entanglement is the essentiality for quantum speedup of certain problems. Except the number of queries to the function, the number of quantum gates in the quantum algorithm in Fig.~3 is also fewer than that in the  quantum algorithm in Fig.~3.

\begin{figure}
\centering
\includegraphics[width=\linewidth]{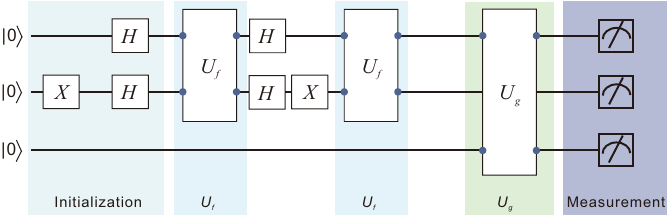}
\caption{(color online). Circuit for the proposed algorithm without entanglement. The information that functions are constant or balanced is stored in the first qubit, while information that they are same or different is stored in the second and third qubits. $U_f$ is applied on the 1-st and 2-nd qubits, and $U_g$ is applied on the 1-st and 3-rd qubits.}
\end{figure}

\section{Experimental realization}

Furthermore, we implement a proof-of-principle experiment to demonstrate the proposed algorithm using the superconducting system \cite{IBMQE}. In our implementation,  we choose two types of balanced function as shown in Fig.~4(a) and Fig.~4(b), and two types of constant function as shown in Fig.~4(c) and Fig.~4(d). Without loss of generality, the  following four cases are considered:

(1) $f=B_1$ and $g=B_1$.

(2) $f=B_1$ and $g=B_2$.

(3) $f=C_1$ and $g=C_1$.

(4) $f=C_1$ and $g=C_2$.\\
where both functions $f$ and $g$ are balanced in case-(1) and case-(2). However, $f=g$ in case-1 and $f \neq g$ in case-(2). Similarly, both functions $f$ and $g$ are constant in case-(3) and case-(4), but $f=g$ in case-3 and $f\neq g$ in case-(4). By substituting the functions in the cases into the circuits in Fig.~(2) and Fig.~(3), we can realize the algorithm with entanglement and without entanglement, respectively.

\begin{figure}[!htbp]
\centering
\includegraphics[width=0.95\linewidth]{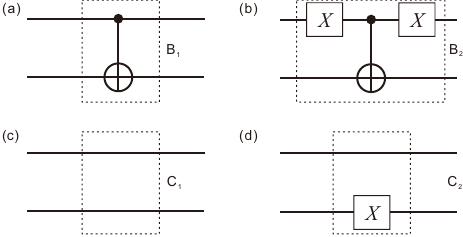}
\caption{(color online). Circuits for different functions. (a) $B_1$ is balanced function, and $B_1(0)=0,B_1(1)=1$. (b) $B_2$ is balanced function, and $B_2(0)=1,B_2(1)=0$.(c) $C_1$ is constant function, and $C_1(0)=C_1(1)=0$. (d) $C_2$ is constant function, and $C_2(0)=C_2(1)=1$.}
\end{figure}

Figure 5(a-d) show both the ideal (red bar) and experimentally obtained (blue bar) probabilities for each outcome when implementing the version of algorithm with entanglement for case-(1-4). We take the Fig.~5(a) as an example to explain the results. Ideally, according to the Table I, with a probability of $50\%$, the output is in ${\rm{|}}100\rangle$, and another $50\%$ probability yields ${\rm{|}}111\rangle$. Whether the measurement result is ${\rm{|}}100\rangle$ or ${\rm{|}}111\rangle$, we can determine that functions $f$ and $g$ are balanced, and $f=g$ according to Table I. To quantify the experimental performance, we use the statistical fidelity ${F  = {\sum\nolimits_{k = 0}^7 \sqrt{p_k^{\text{exp}} p_k^{\text{th}}} }}$ \cite{Carolan} to characterize the overlap between experimental and theoretical values, where ${{p_k^{\text{exp}}}}$ and ${{p_k^{\text{th}}}}$ are the experimental and theoretical output probabilities of the state ${|k\rangle }$, respectively. From the data in Fig.~5, the fidelities are calculated as ${{F _{1}} = 0.891(7)}$, ${{F _{2}} = 0.873(7)}$, ${{F _{3}} = 0.952(7)}$ and ${{F _{4}} = 0.953(7)}$. Thus, the algorithm is announced successful, that is, the version of algorithm with entanglement can solve the task by asking only two queries.

\begin{figure}[htbp]
\centering
\includegraphics[width=\linewidth]{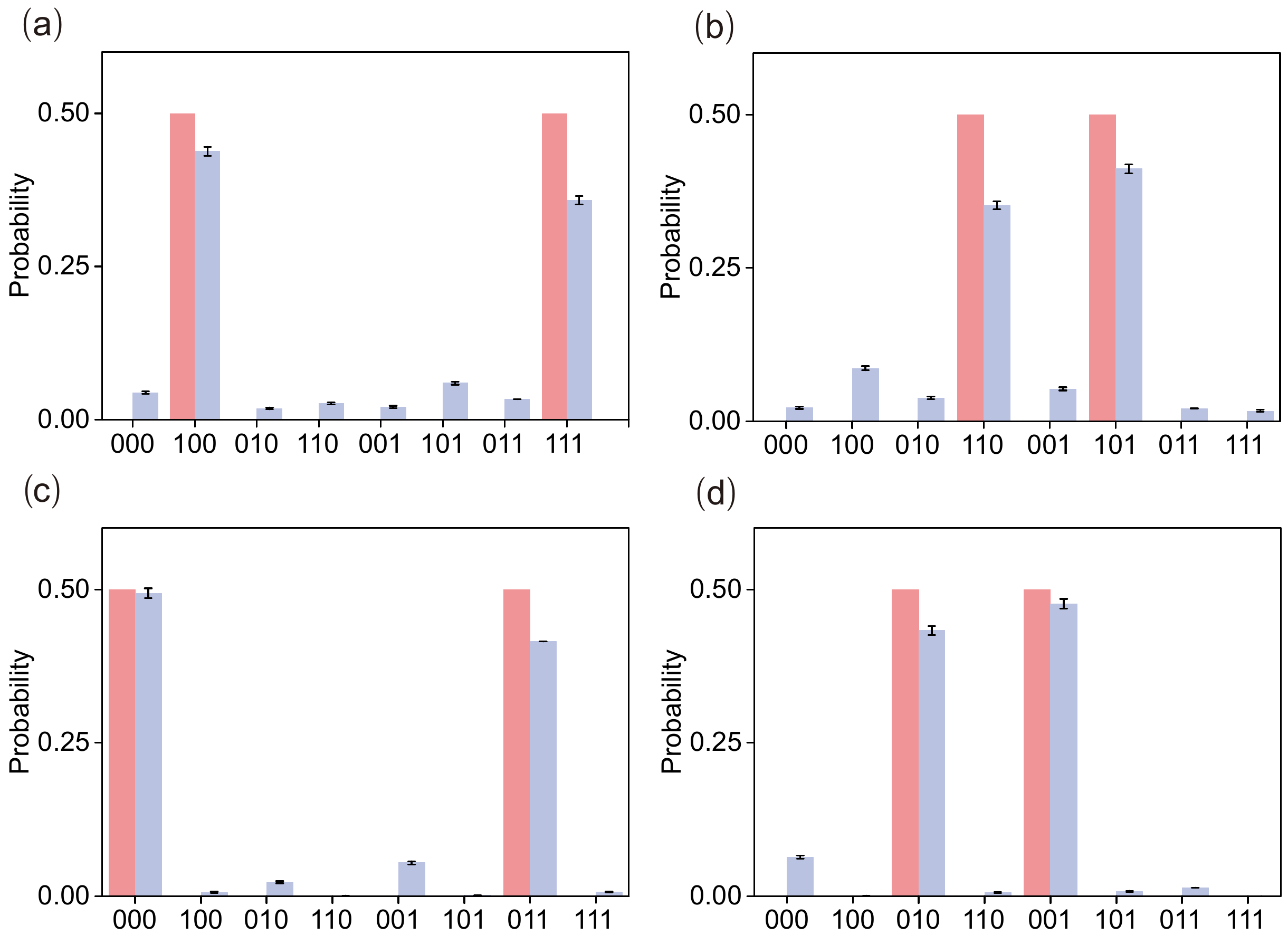}
\caption{(color online). Experimental results. (a-d) are the measurement results of cases-(1-4) by implementing the algorithm with entanglement in Fig.~2. The ideal (red bar) and experimentally obtained (blue bar) probabilities are presented. The error bars denote one standard deviation, deduced from propagated Poissonian counting statistics of the raw detection events.}
\end{figure}

\begin{figure}[htbp]
\centering
\includegraphics[width=\linewidth]{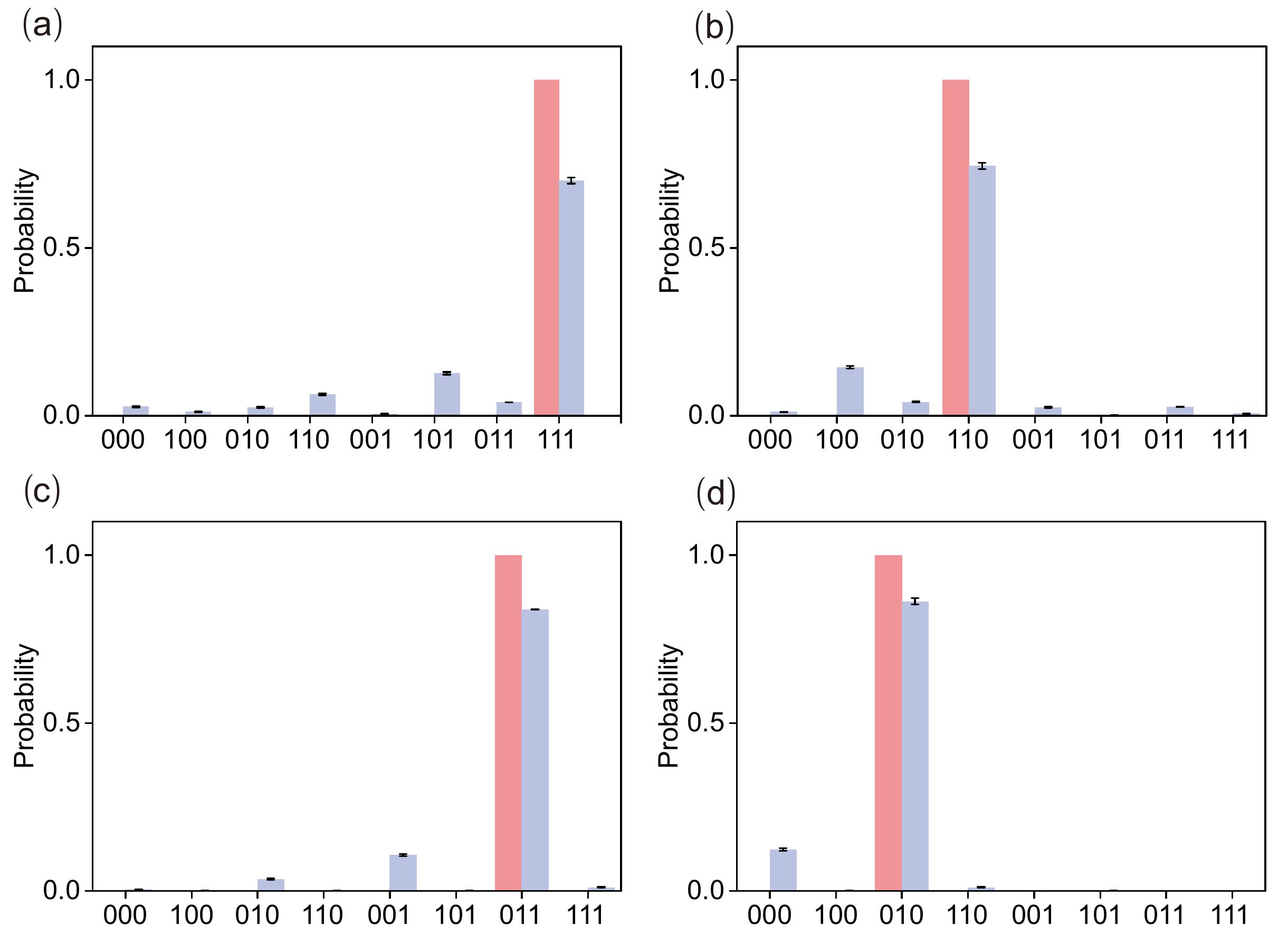}
\caption{(color online). Experimental results. (a-d) are the measurement results of cases-(1-4) by implementing the algorithm without entanglement in Fig.~3.}
\end{figure}

We also implement the version of algorithm without entanglement for case-(1-4) experimentally. Figure 6(a-d) show the measurement results for case-(1-4). From the data in Fig.~6, we can see the experimental results confirm with theoretical prediction. The fidelities of the results for case-(1-4) are  ${{F _{1}} = 0.837(9)}$, ${{F _{2}} = 0.863(10)}$, ${{F _{3}} = 0.916(10)}$ and ${{F _{4}} = 0.929(10)}$. Our experiments show that we can also solve the task by using the version of algorithm without entanglement. However, at least three queries are required. Thus, we have demonstrated that entanglement could help to gain advantage over the quantum algorithm without entanglement.

The imperfections of our experiment mainly arise from the errors in quantum gates and readout. Table II shows the error analysis in our experiment.

\begin{table}[htbp]
\newcommand{\tabincell}[2]{\begin{tabular}{@{}#1@{}}#2\end{tabular}}
\centering
\begin{tabular}{|c |c |c| c| }
\hline
Qubit & 1-st & 2-nd & 3-rd\\
\hline
Gate Error ($10^{-3}$) &1.72&1.46&1.80\\
\hline
Readout Error ($10^{-2}$) &4.20&7.00&1.40\\
\hline
MultiQubit Gate Error ($10^{-2}$) & \tabincell{c}{CNOT$_{12}$\\ 3.17} &\tabincell{c}{CNOT$_{23}$\\ 2.87}&\tabincell{c}{CNOT$_{13}$\\ 2.67}\\
\hline
\end{tabular}
\caption{Error analysis of the superconducting quantum computer. $\text{CNOT}_{ab}$ is the CNOT between $a$ qubit (control qubit) and $b$ qubit (target qubit).}
\end{table}
\section{Conclusions}
In summary, a new quantum algorithm has been proposed and demonstrated to illustrate the essential use of quantum entanglement in getting quantum speedup. It has been shown that when restricted to overall two queries to the functions $f$ and $g$, a classical computer or a quantum computer without entanglement can not compute logical functions, $f(0) \oplus f(1)$ or $f(0) \oplus g(0)$ together. However, a quantum computer having entanglement as a resource can compute these quantities deterministically with one query to each of the functions $f$ and $g$. The algorithm that we demonstrated here, succinctly illustrates the way entanglement can be useful for quantum algorithms in a simple way, which could be used as a prototype in future to develop useful quantum algorithms. Furthermore, our proposed algorithm could be directly used to learning the property of two Boolean functions (testing whether two Boolean functions are the same), and could be easily generalized to test more Boolean functions \cite{Gangopadhyay}.

\textbf{Acknowledgement:} We thank Dintomon Joy for helpful discussions. The authors acknowledge the use of IBM's Quantum Experience for this work. The views expressed are those of the author and do not reflect the official policy or position of IBM or the IBM Quantum Experience team. This work was supported by the National Basic Research Program of China (Grant No. 2013CB338002), National Natural Science Foundation of China(Grants No. 11504430 and No. 61502526).

 (He-Liang Huang and Ashutosh K. Goswami contributed equally to this work.)


\end{document}